\documentclass[aps,preprint]{revtex4}%
\usepackage{amsfonts}
\usepackage{amsmath}
\usepackage{amssymb}
\usepackage{graphicx}%
\begin{document}
\title{{\LARGE ROTATION IN A GENERALISED DIRAC'S UNIVERSE}}
\author{Marcelo Samuel Berman$^{1}$}
\affiliation{$^{1}$Instituto Albert Einstein\ / Latinamerica- Av. Candido Hartmann, 575 -
\ \# 17}
\affiliation{80730-440 - Curitiba - PR - Brazil - email: msberman@institutoalberteinstein.org}
\date{(First Version: 02 April, 2004, Last Version: 26 October, 2008).}

\begin{abstract}
The extension of Dirac's LNH to cover cosmological and fine-structure
\ time-varying "constants", \ and the rotation of the Universe, is here
analysed, including a "derivation" of the angular speed of the present
Universe, and of the inflationary phase. Criticizable points on the present
calculation, are clarified.


\textbf{Resumen:}

Extendiendo la Hip\'{o}tesis de los Grandes N\'{u}meros de Dirac, para cubrir
las "constantes" cosmol\'{o}gica y de estructura fina, y la rotaci\'{o}n del
Universo, las estudiamos, incluyendo una "demonstraci\'{o}n", del c\'{a}lculo
de la velocidad angular del Universo actual, y del inflacion\'{a}rio.Puntos
criticables de \'{e}ste \ art\'{\i}culo, son clarificados \ en seguida.

\end{abstract}
\maketitle

\begin{center}
{\LARGE ROTATION IN A GENERALISED DIRAC'S UNIVERSE}

Marcelo Samuel Berman
\end{center}

\bigskip

{\LARGE \bigskip1. Introduction}

\bigskip We shall study a generalisation of Dirac's LNH\ Universe, with the
introduction of time-varying speed of light, which causes a time-varying
fine-structure "constant", and a possible rotation of the Universe, either for
the present time, or for inflationary periods.

\bigskip

The rotation of the Universe (de Sabbata and Sivaram, 1994; de Sabbata and
Gasperini, 1979) may have been detected experimentally by NASA scientists who
tracked the Pioneer probes, \ finding an anomalous deceleration that affected
the spaceships during the thirty years that they took to leave the Solar
system. This acceleration can be explained through the rotation of the Machian
Universe (Berman, 2007b). Berman (2008) showed that Robertson-Walker's metric
includes a rotational component, plus expansion. Berman (2008d) gave a follow
up to the subject obtaining  the same results. A universal spin has also been
considered by Berman (2008b; 2008c).

\bigskip A time-varying gravitational constant, as well as others, were
conceived by P.A.M. Dirac (1938; 1974), Eddington (1933; 1935; 1939), Barrow
(1990) through his Large Number Hypothesis. Later, Berman supplied the GLNH --
Generalised Large Number Hypothesis (Berman, 1992; 1992a; 1994). This
hypothesis arose from the fact that certain relationships among physical
quantities, revealed extraordinary large numbers of the order \ $10^{40}$\ \ .
Such numbers, instead of being coincidental and far from usual values, were
attributed to time-varying quantities, related to the growing number of
nucleons in the Universe.\ In fact, such number \ \ $N$\ \ , for the present
Universe, is estimated as \ $\left(  10^{40}\right)  ^{2}$\ \ . The number is
"large" because the Universe is "old". At least, this was and still is the
best explanation at our disposal.

\bigskip

The four relations below, represent respectively, the ratios among the scalar
length of the causally related Universe, and the Classical electronic radius;
the ratio between the electrostatic and gravitational forces between a proton
and an electron; the mass of the Universe divided by the mass of a proton or a
nucleon; and a relation involving the cosmological constant and the masses of
neutron and electron.

\bigskip

If we call Hubble's constant \ \ $H$\ \ ; electron's charge and mass
\ $e$\ \ , \ \ $m_{e}$\ \ ; proton's mass \ $m_{p}$\ \ , cosmological constant
\ \ \ $\Lambda$\ \ , speed of light \ $c$\ \ , and Planck's constant
\ $h$\ \ , we have:\ \ 

\bigskip

\bigskip\hfill$\frac{cH^{-1}}{(\frac{e^{2}}{m_{e}c^{2}})}\cong\sqrt{N}$ \ .
\hfill(1)

\bigskip\hfill$\frac{e^{2}}{Gm_{p}m_{e}}\cong\sqrt{N}$ \ . \hfill(2)

\bigskip\hfill$\frac{\rho(cH^{-1})^{3}}{m_{p}}\cong N$ \ \ . \hfill(3)

\bigskip\hfill$ch(m_{p}m_{e}/\Lambda)^{1/2}\cong\sqrt{N}$ \ . \hfill\ (4)

\bigskip

\bigskip We may in general have time-varying speed of light \ $c=c(t)$\ \ ; of
\ $\Lambda=\Lambda(t)$\ \ ; of \ \ $G=G(t)$\ \ ; etc.\ We define the fine
structure "constant" as,

\bigskip\hfill$\alpha\equiv\frac{e^{2}}{\bar{h}\text{ }c(t)}$
\ \ \ \ \ \ \ ,\hfill(5)

\bigskip

and consider \ $\alpha=\alpha(t)$\ \ , because of the time-varying speed of light.

\bigskip

{\LARGE 2. Power-law variations}

\bigskip

\bigskip It is well-known that experimental data should be, by necessity,
interpreted against the background theoretical framework that is satisfiable
from the point of view of the majority of researchers in the field. We shall
choose Robertson-Walker's metric as such background, \ and power-law scale-factors.

\bigskip

One can ask whether the previous Section's constant-variations could be caused
by \ a time-varying speed of light: $c=c(t)$. \ We refer to Berman (2007) for
information on the experimental time variability of \ $\alpha$ \ \ . Gomide (
1976 ) has studied \ \ \ \ $c(t)$\ \ and \ $\alpha$ \ in\ \ such a case, which
was later revived by Barrow (1998 ;\ 1998a ; 1997); Barrow and Magueijo ( 1999
); Albrecht and Magueijo (1998); Bekenstein (1992). This could explain also
Supernovae observations. We refer to their papers for further information. Our
framework now will be an estimate made through Berman's GLNH. We write again
the main formulae for constant deceleration parameters, as derived originally
in the literature by Berman ( 1983 ) and Berman and Gomide ( 1988 ):

\bigskip\hfill$H=\frac{\dot{R}}{R}=\frac{1}{mt}=\frac{1}{1+q}t^{-1}$ . \hfill(6)

\bigskip\hfill$q=-\frac{\ddot{R}R}{\dot{R}^{2}}=m-1$ . \hfill(7)

\bigskip\hfill$R=(mDt)^{1/m}$ \ \ . \ \ \ \ \ ( $m$ $\neq$ 0 )\ ,\hfill(8)

\bigskip\noindent where $R$ is the scale-factor in Robertson-Walker's metric:

\bigskip\hfill$ds^{2}=dt^{2}-\frac{R^{2}(t)}{(1+\frac{kr^{2}}{4})^{2}}%
(dx^{2}+dy^{2}+dz^{2})$ .\hfill(9)

\bigskip

We express now Webb et al's (1999; 2001) experimental result as:

\bigskip\hfill$\left(  \frac{\dot{\alpha}}{\alpha}\right)  _{\exp}\simeq-1.1$
x $10^{-5}$ $H$ $(1+q)$ . \hfill(10)

\bigskip

{\LARGE \bigskip}From (5) we find:

\bigskip\hfill$\frac{\dot{\alpha}}{\alpha}=-\frac{\dot{c}}{c}$ \ . \hfill(11)

\bigskip

Again, we suppose that the speed of light varies with a power law of time:

\bigskip\hfill$c=At^{n}$ \ \ \ \ \ \ ( $A$ = constant ) \ . \hfill(12)

\bigskip

From the above experimental value we find: \ \ 

\bigskip\hfill$n\approx$ $10^{-5}$ . \hfill(13)

\bigskip From (12) and (13) taken care of \ (11), we find:

\bigskip\hfill$\frac{\dot{\alpha}}{\alpha}=-\frac{\dot{c}}{c}=nt^{-1}$ \ .
\hfill(14 )

\bigskip From relations \ \ (1), (2), (3) and (4) we find:

\bigskip

\ \ \ \ \ \ \ \ \ \ \ \ \ \ \ \ \ \ \ \ \ \ \ \ \ \ \ \ \ \ \ \ \ \ \ $N\propto
t^{2+6n}$

\bigskip

\ \ \ \ \ \ \ \ \ \ \ \ \ \ \ \ \ \ \ \ \ \ \ \ \ \ \ \ \ \ \ \ \ \ $G\propto
t^{-1-3n}$

\bigskip

\bigskip
\ \ \ \ \ \ \ \ \ \ \ \ \ \ \ \ \ \ \ \ \ \ \ \ \ \ \ \ \ \ \ \ \ \ \ $\Lambda
\propto t^{-2-4n}$

\bigskip

\bigskip
\ \ \ \ \ \ \ \ \ \ \ \ \ \ \ \ \ \ \ \ \ \ \ \ \ \ \ \ \ \ \ \ \ \ \ $\rho
\propto t^{-1+3n}$

\bigskip

We see that the speed of light varies slowly with the age of the Universe. For
the numerical value (13), we would obtain:

\bigskip\ \ \ \ 

\bigskip\hfill$N\propto t^{2.0001}$ \ , \hfill(15 )

\bigskip\noindent and then:

\bigskip\hfill$G\propto t^{-1.00005}$ \ . \hfill(16)

\bigskip\hfill$\Lambda\propto t^{-2.0001}$ \ \ . \hfill(17)

\bigskip\hfill$\rho\propto t^{-0.99995}$ \ . \hfill(18)

\bigskip

This is our solution, based on Berman's GLNH, itself \ based on Dirac's work
(Dirac, 1938; 1974). A pre-print with a preliminary but incomplete solution
was already prepared by Berman and Trevisan ( 2001; 2001a; 2001b ).

\bigskip

A close analysis shows how the deceleration parameter range is situated. Both
from Lunar laser ranging and Viking radar measurements by Williams et al (
1976 ) and Reasenberg ( 1983 ) we find:

\bigskip\hfill$\frac{\dot{G}}{G}=\sigma H$ \ \ \ \ \ \ with \ $\ \ \ |\sigma
|<0.6$ \ \ .\hfill\hfill

\bigskip

Will ( 1987; 1995 ) and Dickey ( 1994 ) comment that these two kinds of
measurements give the best limit on $|\sigma|.$ Because in our model we have:

\bigskip\hfill$\frac{\dot{G}}{G}$ $\cong1.0$ $t^{-1}\cong1.0$ $H$ $(1+q)$ ,
\hfill(19 )

\bigskip\noindent we find:

\bigskip\hfill$-0.4>q>-1.6$ \ . \hfill(20)

\bigskip

The above result is satisfactory, for a negative "$q$" \ agrees with
Supernovae observations.

\bigskip

We have thus shown that $\frac{\Delta\alpha}{\alpha}$ should really be
negative, for a positive value, could mean a positive deceleration parameter.
As a bonus we found possible laws of variation for $N$, $G$, $\rho$, and
$\Lambda.$ The $\Lambda$--term time variation is also very close and even,
practically indistinguishable, \ from the law of variation $\Lambda$ $\propto
t^{-2}$ \ .

\bigskip

It is clear that in this Section's model, the electric permittivity of the
vacuum, along with its magnetic permeability, and also Planck's constant are
really constant here. We point out again, that in the long run, it will be
only when a Superunification theory becomes available, that our different
models could be discarded, ( hopefully ) but one.

\bigskip

{\large 3. Exponential Inflation}

\bigskip

On remembering that relations (1) and (3) carry the scale-factor of the
causally related Universe, \ \ $cH^{-1}$\ \ , we substitute it by the
exponential radius, \ 

$\bigskip$

$R=R_{0}e^{Ht}$\ \ \ \ \ \ \ \ \ \ \ \ \ \ \ \ \ \ \ . \ \ \ \ \ \ \ \ \ \ \ \ \ \ \ \ \ \ \ \ \ \ \ \ \ \ \ \ \ \ \ \ \ \ \ \ \ \ \ \ \ \ \ \ \ \ \ \ \ \ \ \ \ \ \ \ \ \ \ \ \ \ \ \ \ \ \ \ \ \ \ \ \ \ \ \ \ \ \ \ (21)

\bigskip

With the same arguments above, but, substituting, (12)\ by the following one,

\bigskip

$c=c_{0}e^{\gamma t}$ \ \ \ \ \ \ \ \ \ \ \ \ \ \ \ \ \ \ \ \ , \ \ \ (
$c_{0}$\ , \ $\gamma$\ \ = \ constants)\ \ \ \ \ \ \ \ \ \ \ \ \ \ \ \ \ \ \ \ \ \ \ \ \ \ \ \ \ \ \ \ \ \ \ \ \ \ \ \ \ \ \ \ \ (22)

\bigskip

we would find:

\bigskip

$N\propto e^{\left[  H+2\gamma\right]  \text{ }t}$ \ \ \ \ \ \ \ \ \ \ \ \ \ \ \ \ \ ,

\bigskip

$G\propto e^{-\left[  \frac{H}{2}+\gamma\right]  \text{ }t}$ \ \ \ \ \ \ \ \ \ \ \ \ \ \ ,

\bigskip

$\rho\propto e^{-2\left[  H-\gamma\right]  \text{ }t}$ \ \ \ \ \ \ \ \ \ \ \ \ \ \ \ \ \ ,

\bigskip

and,

\bigskip

$\Lambda\propto e^{-H\text{ }t}$ \ \ \ \ \ \ \ \ \ \ \ \ \ \ \ \ \ \ \ .

\bigskip

It seems reasonable that inflation decreases the energy density, and the
cosmological term while \ $N$\ \ \ grows exponentially; of course, we take
$H>\gamma$\ \ .\ \ 

\bigskip

{\LARGE \bigskip4. Rotation of the Universe}

\bigskip A closely related issue is the possibility of a Universal spin.
Consider the Newtonian definition of angular momentum \ $L$\ \ ,

\bigskip

$L=RMv$ \ \ \ \ \ \ \ \ \ \ \ \ \ \ \ \ \ \ \ \ , \ \ \ \ \ \ \ \ \ \ \ \ \ \ \ \ \ \ \ \ \ \ \ \ \ \ \ \ \ \ \ \ \ \ \ \ \ \ \ \ \ \ \ \ \ \ \ \ \ \ \ \ \ \ \ \ \ \ \ \ \ \ \ \ \ \ \ \ \ \ \ \ \ \ \ \ \ \ \ \ (23)

\bigskip

where, \ $R$\ \ and \ \ $M$\ \ stand for the scale-factor and mass of the Universe.

\bigskip

For Planck's Universe, the obvious dimensional combination of the constants
\ $\ \bar{h}$ \ \ , $c$ \ \ , and \ $G$\ \ \ is,

\bigskip

$L_{Pl}=$ \ $\bar{h}$\ \ \ \ \ \ . \ \ \ \ \ \ \ \ \ \ \ \ \ \ \ \ \ \ \ \ \ \ \ \ \ \ \ \ \ \ \ \ \ \ \ \ \ \ \ \ \ \ \ \ \ \ \ \ \ \ \ \ \ \ \ \ \ \ \ \ \ \ \ \ \ \ \ \ \ \ \ \ \ \ \ \ \ \ \ \ \ \ \ \ \ \ \ \ \ \ \ (24)

\bigskip

From \ (23) and (24), we see that Planck's Universe spin takes a speed
\ $v=c$\ \ . For any other time, we take, then, the spin of the Universe as
given by

\bigskip

\ $L=RMc$ \ \ \ \ \ \ \ \ \ \ \ \ \ \ \ \ \ \ \ \ . \ \ \ \ \ \ \ \ \ \ \ \ \ \ \ \ \ \ \ \ \ \ \ \ \ \ \ \ \ \ \ \ \ \ \ \ \ \ \ \ \ \ \ \ \ \ \ \ \ \ \ \ \ \ \ \ \ \ \ \ \ \ \ \ \ \ \ \ \ \ \ \ \ \ \ (25)

\bigskip

In the first place, we take the known values of the present Universe:

\bigskip

$R\approx10^{28}cm$ \ \ \ \ \ \ \ \ \ \ \ \ \ \ \ \ \ \ \ ,

\bigskip

and,

\bigskip

$M\approx10^{55}grams$ \ \ \ \ \ \ \ \ \ \ \ \ \ ,

\bigskip

so that,

\bigskip

$L=10^{93}cm.gram.cm/s=10^{120}$ $\bar{h}$ . \ \ \ \ \ \ \ \ \ \ \ \ \ \ \ \ \ \ \ \ \ \ \ \ \ \ \ \ \ \ \ \ \ \ \ \ \ \ \ \ \ \ \ \ \ \ \ \ \ \ \ \ \ \ \ \ \ \ \ \ (26)

\bigskip

\bigskip We have thus, another large number, \ 

$\bigskip$

$\frac{L}{\bar{h}}\propto N^{3/2}$\ \ \ \ \ \ \ \ \ \ \ \ \ \ \ . \ \ \ \ \ \ \ \ \ \ \ \ \ \ \ \ \ \ \ \ \ \ \ \ \ \ \ \ \ \ \ \ \ \ \ \ \ \ \ \ \ \ \ \ \ \ \ \ \ \ \ \ \ \ \ \ \ \ \ \ \ \ \ \ \ \ \ \ \ \ \ \ \ \ \ \ \ \ \ \ \ \ \ (27)\ \ \ 

\bigskip

For instance, for the power law, as in standard cosmology, we would have ,

\bigskip

$L\propto t^{3+9n}=t^{3\left(  1+3n\right)  }$
\ \ \ \ \ \ \ \ \ \ \ \ \ \ \ \ . \ \ \ \ \ \ \ \ \ \ \ \ \ \ \ \ \ \ \ \ \ \ \ \ \ \ \ \ \ \ \ \ \ \ \ \ \ \ \ \ \ \ \ \ \ \ \ \ \ \ \ \ \ \ \ \ \ \ \ \ \ \ \ \ \ \ \ \ (28)

\bigskip

For exponential inflation, \ 

$\bigskip$

$L\propto e^{\frac{3}{2}\left[  H+2\gamma\right]  \text{ }t}$%
\ \ \ \ \ \ \ \ \ \ \ \ \ \ . \ \ \ \ \ \ \ \ \ \ \ \ \ \ \ \ \ \ \ \ \ \ \ \ \ \ \ \ \ \ \ \ \ \ \ \ \ \ \ \ \ \ \ \ \ \ \ \ \ \ \ \ \ \ \ \ \ \ \ \ \ \ \ \ \ \ \ \ \ \ \ \ \ \ \ \ \ \ \ \ (29)

\bigskip

\bigskip We now may guess a possible angular speed of the Universe, on the
basis of Dirac's \ LNH. For Planck's Universe, the obvious angular speed would be:

\bigskip

$\omega_{Pl}=\frac{c}{R_{Pl}}\approx2$ x $10^{43}s^{-1}$
\ \ \ \ \ \ \ \ \ \ \ \ \ \ \ \ \ \ \ \ , \ \ \ \ \ \ \ \ \ \ \ \ \ \ \ \ \ \ \ \ \ \ \ \ \ \ \ \ \ \ \ \ \ \ \ \ \ \ \ \ \ \ \ \ \ \ \ \ \ \ (30)

\bigskip

because Planck's Universe is composed of dimensional combinations of the
fundamental constants.

\bigskip

In order to get a time-varying function for the angular speed, we recall
Newtonian angular momentum formula,

\bigskip

$L=R^{2}M\omega$ \ \ \ \ \ \ \ \ \ \ \ \ \ \ \ \ . \ \ \ \ \ \ \ \ \ \ \ \ \ \ \ \ \ \ \ \ \ \ \ \ \ \ \ \ \ \ \ \ \ \ \ \ \ \ \ \ \ \ \ \ \ \ \ \ \ \ \ \ \ \ \ \ \ \ \ \ \ \ \ \ \ \ \ \ \ \ \ \ \ (31)

\bigskip

In the case of power-law $c$\ -- variation\ , we have found, from relation
(27), that,  \ \ $L\propto N^{3/2}$\ \ , but we also saw from (31) that
\ $L\propto\rho R^{5}\omega$\ \ , \bigskip because \ $R=cH^{-1}\propto\sqrt
{N}$\ \ and \ \ $M\propto\rho R^{3}\propto N$\ \ \ \ . 

\bigskip

Then, we find that,

\bigskip

$\omega=\omega_{0}t^{-1+6n}=AR^{-\left(  1-6n\right)  }$
\ \ \ \ \ \ \ \ \ \ \ ( \ $\omega_{0}$\ ,\ $A$\ $=$\ constants )  \ \ \ \ \ . \ \ \ \ \ \ \ \ \ \ \ \ \ \ \ \ \ \ \ (31a)

\bigskip

We are led to admit the following relation:

\bigskip

$\omega\lessapprox\frac{c}{R}$ \ \ \ \ \ \ \ \ \ \ \ \ . \ \ \ \ \ \ \ \ \ \ \ \ \ \ \ \ \ \ \ \ \ \ \ \ \ \ \ \ \ \ \ \ \ \ \ \ \ \ \ \ \ \ \ \ \ \ \ \ \ \ \ \ \ \ \ \ \ \ \ \ \ \ \ \ \ \ \ \ \ \ \ \ \ \ \ \ \ \ \ \ \ \ \ \ \ (32)

\bigskip

For the present Universe, we shall find,

\bigskip

$\omega\lessapprox3$ x $10^{-18}s^{-1}$ \ \ \ \ \ \ \ \ \ \ \ \ . \ \ \ \ \ \ \ \ \ \ \ \ \ \ \ \ \ \ \ \ \ \ \ \ \ \ \ \ \ \ \ \ \ \ \ \ \ \ \ \ \ \ \ \ \ \ \ \ \ \ \ \ \ \ \ \ \ \ \ \ \ \ \ \ \ \ \ \ \ \ (33)

\bigskip

It can be seen that present angular speed is too small to be detected by
present technology.

\bigskip

For the inflationary model, we carry a similar procedure:

\bigskip

$\omega\propto\frac{N^{\frac{3}{2}}}{R^{5}\rho}=e^{\left[  -\frac{9}%
{2}H+\gamma\right]  \text{ }t}$ \ \ \ \ \ \ \ \ \ \ \ \ \ \ \ \ \ \ \ \ . \ \ \ \ \ \ \ \ \ \ \ \ \ \ \ \ \ \ \ \ \ \ \ \ \ \ \ \ \ \ \ \ \ \ \ \ \ \ \ \ \ \ \ \ \ \ \ \ \ \ \ \ \ \ \ \ \ \ \ (34)

\bigskip

\bigskip The condition for a decreasing angular speed in the inflationary
period, is, then, 

\bigskip

$\gamma<\frac{9}{2}H$ \ \ \ \ \ \ \ \ \ \ \ \ . \ \ \ \ \ \ \ \ \ \ \ \ \ \ \ \ \ \ \ \ \ \ \ \ \ \ \ \ \ \ \ \ \ \ \ \ \ \ \ \ \ \ \ \ \ \ \ \ \ \ \ \ \ \ \ \ \ \ \ \ \ \ \ \ \ \ \ \ \ \ \ \ \ \ \ \ \ \ \ \ \ \ \ \ \ \ \ \ \ (35)

{\LARGE 5. Pros and Cons of the present calculations}

\bigskip

Critical appraisals of the above calculations, center on the three following arguments:

\bigskip

\textbf{I - }do the time variations, \ $G(t)$\ \ , \ \ \ $\rho(t)$\ \ \ ,
\ and \ \ $\Lambda(t)$\ \ \ proposed above, violate Einstein's field equations?

\bigskip

\textbf{II -} if Einstein's theory does not apply, which one does? And then,
do the new equations reduce to Einstein's in a proper limit?

\bigskip

\textbf{III -} the identification of \ $R$\ \ with \ \ $cH^{-1}$\ \ is a
non-sense, for the former has no causal meaning while the latter is the radius
of the casually\ connected portion of space associated with Hubble's horizon.
Each one has different time variations.

\bigskip

We now reply:

\bigskip

\textbf{1}$^{st}$\textbf{) \ }Dirac never proposed \ LNH\ \ as part of GRT
(General Relativity Theory), neither do I.

\bigskip

\textbf{2}$^{nd}$\textbf{) }\ Robertson-Walker's metric has been employed in
most (perhaps, all) geometrical theories of Cosmology, \ which do reduce
(i.e., are alternative), or not, to GRT in some limit. Dirac's LNH is a foil
for testing hypotheses, like the theoretical frameworks of scalar-tensor
cosmologies with lambda (see for instance Berman, 2007a). In such theories,
\ our present results may be included (Berman, 2007a).

\bigskip

\textbf{3}$^{rd}$\textbf{) }\ The Robertson-Walker's scale-factor
\ $R(t)$\ \ , is defined \ as an adimensional temporally increasing function,
\ but nothing changes if we calibrate \ \ $R(t)$\ \ in such a way that it be
proportional to \ $cH^{-1}$\ \ . See for instance, the derivation of
Robertson-Walker's metric given in the books by Berman (2007; 2007a). Berman
(1997) may have been the first one to explicit seminally such use. Anyhow, in
the power-laws, \ the case \ $m\approx1$\ \ \ , generates the presently known
accelerating Universes, for it points to a possible slightly negative
deceleration parameter \ $q$\ \ , because in the above models, we have from (7),

\bigskip

$m=q+1$\ \ \ \ \ \ \ \ \ \ \ . \ \ \ \ \ \ \ \ \ \ \ \ \ \ \ \ \ \ \ \ \ \ \ \ \ \ \ \ \ \ \ \ \ \ \ \ \ \ \ \ \ \ \ \ \ \ \ \ \ \ \ \ \ \ \ \ \ \ \ \ \ \ \ \ \ \ \ \ \ \ \ \ \ \ \ \ \ \ \ \ \ \ \ \ \ \ \ \ \ 

\bigskip

\ \ \ \ \ \ \ \ \ \ Then, from relation (8), we have,

\bigskip

$R=(mDt)^{1/m}$ $\approx cH^{-1}=ct$\ \ \ \ \ \ \ \ . \ \ \ \ \ ( $m$
$\approx$ $1$ )\ \ .\hfill

\bigskip

\ \ \ \ \ \ \ \ \ \ We hope to have clarified the former cons, with the latter pros.

\bigskip

{\LARGE 6. Conclusions}

\bigskip

Paraphrasing Dicke (1964; 1964a), it has been shown the many faces of Dirac's
LNH, as many as there are about Mach's Principle. In face of \ modern
Cosmology, the naif theory of Dirac is a foil for theoretical discussion on
the foundations of this branch of Physical theory. The angular speed found by
us, matches results by G\"{o}del (see Adler et al., 1975), Sabbata and
Gasperini (1979), and Berman (2007b, 2008a,b,c).

\bigskip

\bigskip There is a \textit{no-rotation} condition, for \ $n=\frac{1}{6}$\ \ ,
in the power-law solution; likewise, with \ $\gamma=\frac{9}{2}H$\ \ , this is
the \textit{no-rotation} condition of the inflationary angular speed formula.
However, these cases are foreign to the idea of a weak time-varying formula
for the fine-structure "constant".

{\LARGE Acknowledgements}

\bigskip

The author thanks his intellectual mentors, Fernando de Mello Gomide and the
late M. M. Som, and also to Marcelo Fermann Guimar\~{a}es, Nelson Suga, Mauro
Tonasse, Antonio F. da F. Teixeira, and for the encouragement by Albert, Paula
and Geni.

\bigskip

\bigskip\bigskip{\Large References}

\bigskip Adler, R.J.; Bazin, M.; Schiffer, M. (1975) - \ \textit{Introduction
to General Relativity, }2$^{nd}$ Edition, McGraw-Hill, New York.

Albrecht,A.; Magueijo, J.(1998) \textit{A time varying speed of light as a
solution to cosmological puzzles}-preprint.

Bahcall, J.N. ; Schmidt, M. (1967) - Physical Review Letters \textbf{19}, 1294.

Barrow, J.D. (1990) - in \textit{Modern Cosmology in Retrospect}, ed. by B.
Bertotti, R.Balbinot, S.Bergia and A.Messina. CUP, Cambridge.

Barrow, J.D. (1997) - \ \ \textit{Varying G and Other Constants}\ , Los Alamos
Archives http://arxiv.org/abs/gr-qc/9711084 v1 27/nov/1997.

Barrow, J.D. (1998) - in \textit{Particle Cosmology, }Proceedings RESCEU
Symposium on Particle Cosmology, Tokyo, Nov 10-13, 1997, ed. K. Sato, T.
Yanagida, and T. Shiromizu, Universal Academic Press, Tokyo, pp. 221-236.

Barrow, J.D. (1998a) - Cosmologies with Varying Light Speed, Los Alamos
Archives http://arxiv.org/abs/Astro-Ph/9811022 v1 .

Barrow, J.D. ; Magueijo, J. (1999) - Solving the Flatness and Quasi-Flatness
Problems in Brans-Dicke Cosmologies with varying Light Speed. Class. Q.
Gravity, 16, 1435-54.

\bigskip Bekenstein, J.D. (1982) - Physical Review \textbf{D25}, 1527.

\begin{description}
\item Berman,M.S. (1983) - \textit{Nuovo Cimento} \textbf{74B}, 182.

\item Berman,M.S. (1992) - \textit{International Journal of Theoretical
Physics} \textbf{31}, 1447.
\end{description}

Berman,M.S. (1992a) - \textit{International Journal of Theoretical Physics}
\textbf{31}, 1217-19.

\bigskip Berman,M. S. (1994) - \textit{Astrophys. Space Science,}
\textbf{215}, 135-136.

Berman,M.S. (1996) - International Journal of Theoretical Physics \textbf{35}, 1789.

Berman,M.S. (2007) - \textit{Introduction to General Relativity and the
Cosmological Constant Problem. }Nova Science, New York.

Berman,M.S. (2007a) - \textit{Introduction to General Relativistic and Scalar
Tensor Cosmologies. }Nova Science, New York.

Berman,M.S. (2007b) - \textit{The Pioneer Anomaly and a Machian Universe,
}Astrophysics and Space Science, \textbf{312}, 275.

Berman,M.S. (2008) - \textit{A General Relativistic Rotating Evolutionary
Universe, }Astrophysics and Space Science, \textbf{314}, 319-321. Preliminary
version, Los Alamos Archives, http://arxiv.org/abs/physics/0712.0821 [physics.gen-ph].

Berman,M.S. (2008b) - \textit{Shear and Vorticity in a Combined
Einstein-Cartan-Brans-Dicke Inflationary Lambda Universe, }Astrophysics and
Space Science, \textbf{314, }79-82. For a preliminary report, see Los Alamos
Archives, http://arxiv.org/abs/physics/0607005

Berman,M.S. (2008c) - \textit{A Primer in Black Holes, Mach's Principle and
Gravitational Energy, }Nova Science, New York.

Berman,M.S. (2008d) - \textit{A General Relativistic Rotating Evolutionary
Universe - Part II, }Astrophysics and Space Science, \textbf{315},367-369.
Posted in Los Alamos Archives, http://arxiv.org/abs/physics/0801.1954 [physics.gen-ph].

Berman,M.S.; Gomide,F.M. (1988) - GRG, \textbf{20}. 191-198.

\begin{description}
\item Berman,M.S.; Trevisan, L.A. (2001)\ - Los Alamos Archives http://arxiv.org/abs/gr-qc/0112011

\item Berman,M.S.; Trevisan, L.A. (2001a) - Los Alamos Archives http://arxiv.org/abs/gr-qc/0111102

\item Berman,M.S.; Trevisan, L.A. (2001b) - Los Alamos Archives http://arxiv.org/abs/gr-qc/0111101

\item de Sabbata, V.; Gasperini, M. (1979) - \textit{Lettere al Nuovo Cimento,
}\textbf{25, }489.

\item de Sabbata, V.; Sivaram, C. (1994) - \textit{Spin and Torsion in
Gravitation, }World Scientific, Singapore.

\item Dicke, R.H. (1964) - \textit{The many faces of Mach, }in
\ \textit{Gravitation and Relativity, }W.A. Benjamin Inc. New York.

\item Dicke, R.H. (1964a) - \textit{The significance for the solar system of
time-varying gravitation, }in \ \textit{Gravitation and Relativity, }W.A.
Benjamin Inc. New York.

\item Dickey, G.O.; et.al. (1994) - Science, \textbf{265}, 482.

\item Dirac, P.A.M. (1938) - \textit{Proceedings of the Royal Society
}\textbf{165} A, 199.

\item Dirac, P.A.M. (1974) - \textit{Proceedings of the Royal Society
}\textbf{A338,} 439.
\end{description}

\bigskip\bigskip Eddington, A.S. (1933) - \textit{Expanding Universe, }CUP, Cambridge.

\bigskip Eddington, A.S. (1935) - \textit{New Pathways in Science}. Cambridge
University Press, Cambridge.

Eddington, A.S. (1939) - Sci. Progress, London, \textbf{34}, 225.

Gomide, F.M. (1976) - \textit{Lett. Nuovo Cimento} \textbf{15}, 515.

Moffat, J.W. (1993) - International Journal of Modern Physics, \textbf{D2}, 351.

Reasenberg, R.D. (1983) - Philosophical Trans. Royal Society, (London),
\textbf{310}, 227.

Sabbata, de V.; Gasperini, M. (1979) - \textit{Lettere al Nuovo Cimento,
}\textbf{25}, 489.

Schwarzschild, B. (2001)- Physics Today, \textbf{54} (7), 16.

Schmidt, B.P. (1998) \ - Ap.J., \textbf{507}, 46-63.

\begin{description}
\item Webb,J.K; et al. (1999)--Phys. Rev. Lett. \textbf{82}, 884.

\item Webb,J.K; et al.(2001)-- Phys Rev. Lett. \textbf{87,} 091301.
\end{description}

Will, C. (1987) - in \textit{300 Years of Gravitation}, ed. by W.Israel and
S.Hawking, CUP, Cambridge.

Will, C. (1995) - in \textit{General Relativity - Proceedings of the 46}%
$^{th}$ \textit{Scotish Universities Summer School in Physics, }ed. by G.Hall
and J.R.Pulhan, IOP/SUSSP Bristol.

Williams, J.G. et al (1976) - Physical Review Letters, \textbf{36}, 551.

\end{document}